\documentstyle[prl,aps,epsf,floats]{revtex}
\begin{document}
\draft
\twocolumn[\hsize\textwidth\columnwidth\hsize\csname @twocolumnfalse\endcsname
\title{Mean-field study of the interplay between antiferromagnetism
       and d-wave superconductivity}
\author{Bumsoo Kyung}
\address{D\'{e}partement de physique and Centre de recherche 
sur les propri\'{e}t\'{e}s \'{e}lectroniques de mat\'{e}riaux avanc\'{e}s. \\
Universit\'{e} de Sherbrooke, Sherbrooke, Qu\'{e}bec, Canada J1K 2R1}
\date{April 17, 2000}
\maketitle
\begin{abstract}

   The interplay between antiferromagnetism and d-wave superconductivity
is studied in a mean-field approximation for 
a generic microscopic Hamiltonian with short-range repulsion and 
near-neighbor attraction.  
In the presence of competing microscopic interactions, the phase boundaries  
of antiferromagnetic and superconducting states
are significantly modified in some region of the doping-temperature
plane.
The transition between superconductivity and antiferromagnetism
occurs through a phase where both order parameters coexist with a third, 
dynamically generated, spin-triplet amplitude. 
This dynamical generation of a new order parameter is not restricted  
to a system with 
antiferromagnetism and d-wave superconductivity,
but is a generic feature for fermionic systems.
The dynamically generated
spin-triplet order parameter is found to be robust
to variations in the mean-field Hamiltonian.
\end{abstract}
\vskip2pc]
\narrowtext

   The close interplay 
between antiferromagnetism (AF) and superconductivity (SC) has  
been of great interest both to experimentalists and theoretical physicists 
for many years. 
The competition of two different order parameters has been observed 
in various systems such as organic  
superconductors,\cite{Jerome:1982} ternary compounds,\cite{Maple:1982} 
heavy fermion compounds\cite{Broholm:1987} as well as high temperature 
superconductors\cite{Bednorz:1986} in the latest.
For some compounds, such as the cuprates, the AF and SC phases are 
close to each other but separated, while for other  
compounds such as organic superconductors, ternary compounds and 
heavy fermion compounds the two phases often touch and sometimes even coexist
in the phase diagram.
This 
competition problem of two different orders was extensively studied in  
a mean-field approximation
first for s-wave superconductors\cite{Machida:1981} and  
later for anisotropic superconductors\cite{Kato:1987,Kato:1988}. 
Recently the same problem including  
the fluctuation effect of AF and SC 
was considered by 
Onoda and Imada\cite{Onoda:19991,Onoda:19992} in the framework of  
self-consistent renormalization theory.

   A few years ago Zhang and his coworkers\cite{Demler:1995,Zhang:1997} 
proposed that the SC and AF phases, particularly, of the high 
$T_{c}$ cuprates might be related to each other in a unified 
theory based on SO(5) symmetry.
In this SO(5) theory, the SC components of the five dimensional superspin 
vector are transformed to the AF components by, so called, 
$\pi$ operators. These operators carry charge 2 and spin 1, and total 
momentum $Q=(\pi,\pi)$ in two dimensions.
Further the above authors showed that a sharp resonance of 41 meV 
observed in inelastic neutron scattering 
experiments\cite{Mook:1993} might be a direct consequence of the existence 
of $\pi$ operators. 
To the author's knowledge, the possibility of spin-triplet  
amplitude with total momentum $Q$ 
often called $\pi$-triplet in the literature,
was first found by Psaltakis and 
Fenton\cite{Psaltakis:1983} in their mean-field study of
the competition  problem of AF and SC.
These authors showed that when a spin-density wave (SDW) coexists with 
spin-singlet Cooper-pair superconductivity among the same electrons,
a non-zero spin-triplet amplitude must appear. 
Recently a similar problem was studied by Murakami and 
Fukuyama\cite{Murakami:1998,Murakami:1999} by invoking the g-ology diagrams 
in one-dimensional electron system. 
But in the above two works the authors did not study 
the condition of appearance of the spin-triplet 
amplitude and its robustness.
Furthermore in previous mean-field studies 
the manner how the AF-only and SC-only 
phase boundaries are modified into new phase boundaries  
in the presence of two competing interactions,
was not well investigated. 
Thus the purpose of this paper is twofold: First, detailed 
phase diagrams are presented in the doping ($x=1-n$)-temperature ($T$) 
plane in the presence of competing interactions
and second the condition of appearance of the spin-triplet 
order parameter and its robustness are clarified.

   In this paper we study a phenomenological mean-field Hamiltonian 
where antiferromagnetism and 
d-wave superconductivity are built in on equal footing. 
\begin{eqnarray}
 H &=& \sum_{\vec{k},\sigma}(\xi_{\vec{k}}-\mu)c^{+}_{\vec{k},\sigma}
      c_{\vec{k},\sigma}
   +U\sum_{i}(c^{+}_{i,\uparrow}  c_{i,\uparrow}
              \langle c^{+}_{i,\downarrow}c_{i,\downarrow} \rangle
             +\mbox{H.C.})
                                             \nonumber  \\
   &-& V\sum_{i}
        ( \Delta^{+}_{d,i}\langle \Delta_{d,i} \rangle 
             +\mbox{H.C.})
       -W\sum_{i}
        ( \Delta^{+}_{t,i}\langle \Delta_{t,i} \rangle
             +\mbox{H.C.}) \; .
                                             \nonumber  \\
                                                           \label{eq1}
\end{eqnarray}
A possible 
spin-triplet pair amplitude $\langle \Delta_{t,i} \rangle $ is also    
considered in general  
for self-consistency and 
its significance will become clear later.
In particular, we will see that it is dynamically generated, even when
$W=0$.
The singlet $\Delta_{d,i}$ 
and triplet $\Delta_{t,i}$ pair destruction operators are defined as 
\begin{eqnarray}
  \Delta_{d,i}
  & = & \frac{1}{2}\sum_{\delta}g(\delta)
        (c_{i+\delta,\uparrow}        c_{i,\downarrow}
        -c_{i+\delta,\downarrow}      c_{i,\uparrow})
                                             \nonumber  \\
  \Delta_{t,i}
 & = & \frac{1}{2}\sum_{\delta}g(\delta)
        (c_{i+\delta,\uparrow}        c_{i,\downarrow}
        +c_{i+\delta,\downarrow}      c_{i,\uparrow}) \; ,
                                                           \label{eq2}
\end{eqnarray}
and the structure factor $g(\delta)$ is chosen to have a d-wave-like form 
such as 
\begin{eqnarray}
 g(\delta) = \left\{ \begin{array}{lll}
                          1/2     &  \mbox{if $\delta=(\pm 1,0)$} , \\
                         -1/2     &  \mbox{if $\delta=(0,\pm 1)$} , \\
                           0      &  \mbox{if otherwise} .
                       \end{array}
               \right.  
                                                           \label{eq3}
\end{eqnarray}
Later it will be also clear why the same structure factor $g(\delta)$ is 
introduced in the spin-triplet term.
The strength of AF, SC, and spin-triplet pair interactions is governed by 
$U$, $V$, and $W$, respectively, which are all positive in our study.
Essentially the same form of the mean-field Hamiltonian was previously  
studied by 
Psaltakis and Fenton,\cite{Psaltakis:1983} and 
Kato and Machida,\cite{Kato:1987} and 
Murakami and Fukuyama\cite{Murakami:1998} except for some slight
differences.
$\xi_{\vec{k}}$ is a tight binding energy dispersion given as
$\xi_{\vec{k}}=-2t(\cos k_{x}+\cos k_{y})
               -4t'\cos k_{x}\cos k_{y}$, where 
$t$ and $t'$ are hopping constants for nearest neighbors and next 
nearest neighbors. And $\mu$ is the chemical potential which controls
the particle density $n$.

   In a mean-field approximation, three different order parameters 
$m$, $s$, and $t$ corresponding to the three different interactions  
may be defined 
in the following way:
\begin{eqnarray}
& & \langle c^{+}_{i,\sigma}c_{i,\sigma} \rangle = 
    \langle n_{i,\sigma} \rangle = \frac{n}{2}+\sigma m \cos(\vec{Q} \cdot
                                                          \vec{R}_{i})
                                             \nonumber  \\
& & \frac{1}{2}\sum_{\delta}g(\delta)
   \langle c_{i+\delta,\uparrow}    c_{i,\downarrow}
   -c_{i+\delta,\downarrow}  c_{i,\uparrow} \rangle = s 
                                             \nonumber  \\
& & \frac{1}{2}\sum_{\delta}g(\delta)
   \langle c_{i+\delta,\uparrow}    c_{i,\downarrow}
   +c_{i+\delta,\downarrow}  c_{i,\uparrow} \rangle = t \cos(\vec{Q} \cdot 
                                              \vec{R}_{i}) \; ,
                                                           \label{eq4}
\end{eqnarray}
where $\vec{Q}$ is the (commensurate) antiferromagnetic wave vector
equal to $(\pi,\pi)$ in two dimensions.
Now the mean-field Hamiltonian $H_{MF}$ is quadratic in the  
original electron operator and 
in terms of a new four component field operator $\psi_{\vec{k}}$
it becomes bilinear 
\begin{eqnarray}
H_{MF}=\sum^{'}_{\vec{k}}\psi^{+}_{\vec{k}}M_{\vec{k}}\psi_{\vec{k}}
       +E_{0} \; ,
                                                           \label{eq5}
\end{eqnarray}
where 
\begin{eqnarray}
\psi^{+}_{\vec{k}}=(c^{+}_{\vec{k},\uparrow},
                    c_{-\vec{k},\downarrow},
                    c^{+}_{\vec{k}+\vec{Q},\uparrow},
                    c_{-\vec{k}-\vec{Q},\downarrow})  \; .
                                                           \label{eq6}
\end{eqnarray}
The matrix $M_{\vec{k}}$ is given as 
\begin{eqnarray}
M_{\vec{k}}= \left(  \begin{array}{cccc}
 \varepsilon_{\vec{k}} & Vs\phi(\vec{k}) & -Um & Wt\phi(\vec{k}) \\
 Vs\phi(\vec{k}) & -\varepsilon_{\vec{k}} & -Wt\phi(\vec{k}) & -Um \\
 -Um & -Wt\phi(\vec{k}) & \varepsilon_{\vec{k}+\vec{Q}} & -Vs\phi(\vec{k}) \\
 Wt\phi(\vec{k}) & -Um & -Vs\phi(\vec{k}) & -\varepsilon_{\vec{k}+\vec{Q}} 
           \end{array}  
                                                           \label{eq7}
   \right)   \; ,
\end{eqnarray}
where 
\begin{eqnarray}
\varepsilon_{\vec{k}}=\xi_{\vec{k}}-\mu  \; 
                                                           \label{eq8}
\end{eqnarray}
and $\phi(\vec{k})=
\cos k_{x}-\cos k_{y}$
is the Fourier transform of $g(\delta)$.
The prime symbol on the summation requires the summation of wave vectors
in {\em half} of the first Brillouin zone,
in order to take into account the doubling of a magnetic unit cell
in the presence of (commensurate) AF order.
The constant energy shift $E_{0}$ depending on $m$, $s$, $t$, and $\mu$
is given as
\begin{eqnarray}
E_{o}=N(Um^{2}+Vs^{2}+Wt^{2}-\mu)  \; ,
                                                           \label{eq9}
\end{eqnarray}
where $N$ is the total number of lattice sites.
The energy eigenvalues of $M_{\vec{k}}$ yield
four energy dispersions $\pm E_{\pm}(\vec{k})$,
\begin{eqnarray}
E_{\pm}(\vec{k})  = & [ &(\varepsilon^2_{\vec{k}}
        +\varepsilon^2_{\vec{k}+\vec{Q}})/2
        +(Um)^2+(Vs\phi(\vec{k}))^2
                                             \nonumber  \\
        &+&(Wt\phi(\vec{k}))^2 \pm g(\vec{k}) \;\;
             ]^{1/2} \; ,
                                                           \label{eq10}
\end{eqnarray}
where $g(\vec{k})$ is given as 
\begin{eqnarray}
g(\vec{k})=&[& (\varepsilon^2_{\vec{k}}-\varepsilon^2_{\vec{k}+\vec{Q}})^2/4
   +(\varepsilon_{\vec{k}}-\varepsilon_{\vec{k}+\vec{Q}})^2
    (Wt\phi(\vec{k}))^2
                                             \nonumber  \\
   &+&((\varepsilon_{\vec{k}}+\varepsilon_{\vec{k}+\vec{Q}})(Um)
    +2(Vs\phi(\vec{k}))(Wt\phi(\vec{k}))
    )^2 \;\;
       ]^{1/2} \; .
                                             \nonumber  \\
                                                           \label{eq11}
\end{eqnarray}
When $s=t=0$ or $m=t=0$, the energy eigenvalue reduces to that 
of SDW or BCS approximation of the corresponding interactions.

   The free energy is easily obtained either from the trace formula 
or from the Feynman theorem
\begin{eqnarray}
F=-2T\sum^{'}_{\vec{k}}\sum_{\alpha=\pm}\log 
               (2\cosh \frac{E_{\alpha}(\vec{k})}{2T})+E_{o}
                                               \; .
                                                           \label{eq12}
\end{eqnarray}
Now the three mean-field equations are obtained by the stationary 
condition of $F$ with respect to the corresponding order parameters
$\frac{\partial F}{\partial m}
=\frac{\partial F}{\partial s}
=\frac{\partial F}{\partial t}=0$ and 
one more unknown constant $\mu$ is determined by the thermodynamic 
relation $n=-\frac{\partial F}{\partial \mu}$.
The resulting four equations are 
\begin{eqnarray}
& m & =  \frac{1}{2N}\sum^{'}_{\vec{k}}\sum_{\alpha=\pm}
     \Bigl\{
           (Um)
           + \alpha\frac
 {(\varepsilon_{\vec{k}}+\varepsilon_{\vec{k}+\vec{Q}})}{2g(\vec{k})}
           [(\varepsilon_{\vec{k}}+\varepsilon_{\vec{k}+\vec{Q}})(Um)
                                             \nonumber  \\
        & & +2(Vs\phi(\vec{k}))(Wt\phi(\vec{k}))]
     \Bigr\}
     \frac{1}{E_{\alpha}}(\vec{k})\tanh(\frac{\beta E_{\alpha}(\vec{k})}{2})
                                             \nonumber  \\
                                              \\
                                                           \label{eq13}
& s & =  \frac{1}{2N}\sum^{'}_{\vec{k}}\sum_{\alpha=\pm}\phi(\vec{k})
     \Bigl\{
           (Vs\phi(\vec{k}))
           + \alpha\frac
           {(Wt\phi(\vec{k}))}{g(\vec{k})}
           [(\varepsilon_{\vec{k}}+\varepsilon_{\vec{k}+\vec{Q}})
                                             \nonumber  \\
        & & (Um)+2(Vs\phi(\vec{k}))(Wt\phi(\vec{k}))]
     \Bigr\}
     \frac{1}{E_{\alpha}}(\vec{k})\tanh(\frac{\beta E_{\alpha}(\vec{k})}{2})
                                             \nonumber  \\
                                               \\
                                                           \label{eq14}
& t & =  \frac{1}{2N}\sum^{'}_{\vec{k}}\sum_{\alpha=\pm}\phi(\vec{k})
     \Bigl\{
           (Wt\phi(\vec{k}))
           + \alpha\frac
           {(Wt\phi(\vec{k}))
 (\varepsilon_{\vec{k}}-\varepsilon_{\vec{k}+\vec{Q}})^2}{2g(\vec{k})}
                                             \nonumber  \\
        & &  + \alpha\frac
           {(Vs\phi(\vec{k}))}{g(\vec{k})}
           [(\varepsilon_{\vec{k}}+\varepsilon_{\vec{k}+\vec{Q}})(Um)
                                             \nonumber  \\
        & & +2(Vs\phi(\vec{k}))(Wt\phi(\vec{k}))]
     \Bigr\}
     \frac{1}{E_{\alpha}}(\vec{k})\tanh(\frac{\beta E_{\alpha}(\vec{k})}{2})
                                             \nonumber  \\
                                               \\
                                                           \label{eq15}
& n & =  1-\frac{1}{2N}\sum^{'}_{\vec{k}}\sum_{\alpha=\pm}
     \Bigl\{
           (\varepsilon_{\vec{k}}+\varepsilon_{\vec{k}+\vec{Q}})
                                             \nonumber  \\
       & &   + \alpha\frac
           {
           (\varepsilon_{\vec{k}}+\varepsilon_{\vec{k}+\vec{Q}})
 (\varepsilon_{\vec{k}}-\varepsilon_{\vec{k}+\vec{Q}})^2}{2g(\vec{k})}
                                             \nonumber  \\
        & &  + \alpha\frac
           {2(Um)}{g(\vec{k})}
           [(\varepsilon_{\vec{k}}+\varepsilon_{\vec{k}+\vec{Q}})(Um)
                                             \nonumber  \\
        & & +2(Vs\phi(\vec{k}))(Wt\phi(\vec{k}))]
     \Bigr\}
     \frac{1}{E_{\alpha}}(\vec{k})\tanh(\frac{\beta E_{\alpha}(\vec{k})}{2})
                                \; .            
                                             \nonumber  \\
                                                           \label{eq16}
\end{eqnarray}
Although 
essentially the same form of the mean-field Hamiltonian was 
studied by 
Psaltakis and Fenton,\cite{Psaltakis:1983} and 
Murakami and Fukuyama\cite{Murakami:1998},
the final mean-field equations are somewhat different.
In Ref.\cite{Psaltakis:1983} the Fermi surface is divided into two regions.
In region one the nesting condition 
$\varepsilon_{\vec{k}}=-\varepsilon_{\vec{k}+\vec{Q}}$ 
is satisfied and in the other region it is not satisfied.
Thus the mean-field equations are different in different regions.
Even after taking this into account there are still some discrepancies. 
In Ref.\cite{Murakami:1998} apparently there are several erroneous 
terms in the mean-field equations.
As a double check for the correctness of the above mean-field equations,
we also computed the transformation matrix relating the original $c$ operator
to the eigen-operator $\gamma$ and with it obtained the same 
mean-field equations. All the components of the matrix are given 
in the Appendix.

   Before plunging into numerical calculations, we point out several 
distinctive features inferred from the structure of 
the mean-field equations itself.
First, a given order parameter is affected by the other ones
in a more direct manner than for the case where only two order 
parameters, AF and SC,\cite{Kato:1987} are considered. For the latter case
the influence from the other order parameters comes indirectly through 
$E_{\alpha}(\vec{k})$ and $g(\vec{k})$.
Second, in spite of non-vanishing driving potentials for the three different 
order parameters, only a state with a single order parameter 
or with three order parameters together is possible.
For the former case the proof is obvious.
Suppose two of the three order parameters, for example, $m$ and $s$,
are non-vanishing and then examine whether the third one ($t$)
can vanish or not.
Since the equations have the same structure with respect to the three 
order parameters, this particular choice suffices for general purpose.
The right hand of the $t$ equation, Eq. 15, has a term 
which is independent of
$t$,
\begin{eqnarray}
(Vs\phi(\vec{k}))
(\varepsilon_{\vec{k}}+\varepsilon_{\vec{k}+\vec{Q}})(Um) \; .
                                                           \label{eq17}
\end{eqnarray}
Thus unless this term is exactly canceled by the combination of the 
other terms, the third order parameter $t$ does not vanish. But perfect 
cancellation is impossible because the term Eq. 17 cannot 
be written as a linear combination of the other terms in Eq. 15. 
Third, despite the vanishing driving potential for one of the order 
parameters, for example, $W=0$,
the corresponding order parameter $t$ (in this case) can be {\em dynamically} 
generated.
Since the terms containing $t$ enter with the combination of  
$Wt\phi(\vec{k})$
in the right hand of Eq. 15, 
only Eq.\ref{eq17} survives for $W=0$.
As long as both AF and SC order parameters have non-vanishing amplitude,
namely, they coexist,
$t$ also has non-vanishing amplitude even with $W=0$. 

    This last feature was first discussed 
by Psaltakis and 
Fenton\cite{Psaltakis:1983}. 
They argued that when AF and SC order parameters 
$\langle c^{+}_{\vec{k},\uparrow}c_{\vec{k}+\vec{Q},\uparrow} \rangle$,
$\langle c_{\vec{k},\uparrow}c_{-\vec{k},\downarrow} \rangle$
have non-vanishing value at the same time,
two electrons
$c_{\vec{k}+\vec{Q},\uparrow}$,
$c_{-\vec{k},\downarrow}$
are coupled to the same electron
$c^{+}_{\vec{k},\uparrow}$ through the Bogoliubov transformation,
thus they 
are not generally independent but instead may form another order parameter
$\langle c_{\vec{k}+\vec{Q},\uparrow}
c_{-\vec{k},\downarrow} \rangle$.
Eq.\ref{eq17} shows in which circumstances this may happen.
The d-wave factor $\phi(\vec{k})$ in front of the curly bracket in 
Eq. 15 comes from the spin-triplet term and 
$\phi(\vec{k})$ in Eq.\ref{eq17} from the spin-singlet term, thus 
the same symmetry factors, or at least those which are not orthogonal 
to each other,
are required to obtain a non-vanishing third order parameter.
At half-filling for a nearest neighbor model ($t'=0$) where 
$\varepsilon_{\vec{k}}+\varepsilon_{\vec{k}+\vec{Q}}=0$,
however,
the dynamical generation of another order parameter does not happen.
Since the equations again have the same structure with respect to the three 
order parameters, $m$ ($s$) can be dynamically generated  with 
$U=0$ ($V=0$) as well.
In this paper we argue that 
this dynamical generation of third order parameter 
in the coexistence state
of two orders is not restricted only to the case with AF and d-wave SC orders,
but is a generic feature for fermionic systems.
With a proper choice of the symmetry factor, 
a third order parameter can be dynamically generated without the 
corresponding driving potential. 
For example,
for the case of SDW and p-wave SC, the third order parameter is 
$\sum_{\vec{k}} \phi(\vec{k}) \langle c_{\vec{k},\uparrow}
                                      c_{\vec{k}+\vec{Q},\downarrow} \rangle$,
where $\phi(\vec{k})$ is a p-wave form factor such as 
$\sin k_{x}$ or $\sin k_{y}$.
This may be called $\pi$-singlet in parallel to 
the $\pi$-triplet for the case of SDW and d-wave SC.
In SO(5) theory, the $\pi$-triplet amplitude is not an order parameter.
It is the generator of infinitesimal rotations between AF and SC 
order parameters. In the mixed AF+SC+$\pi$-triplet phase that is 
discussed here, the $\pi$-triplet itself acquires a non-zero expectation
value.

    First we start with presenting phase diagrams of the nearest neighbor
model ($t'=0$).   
In this paper,    
$t$, lattice spacing, $\hbar$, and $k_{B}$ are set to be unity, and 
$U$ is fixed at half of the bandwidth, $4$.
With varying $V$, the shape of the phase diagram changes significantly. 
In the present study we restrict ourselves to a particular case where 
at half filling the ground state is an antiferromagnetic insulator,
which might be more relevant especially for high $T_{c}$ cuprates.
This requirement dictates that $V$ is less than $U=4$ and thus we consider 
two different values of $V$, $3$ and $1.5$.
In Fig.\ref{fig1} and in the following figures, 
the dotted and dashed curves denote the phase boundary of AF and 
SC without the driving interaction term for the other order 
parameter. For instance, the dotted curve in Fig.\ref{fig1}-(a) is 
the phase boundary of AF state with $V=0$ and the dashed one of 
SC state with $U=0$.
Fig.\ref{fig1} clearly shows how the AF-only (dotted curve) and SC-only 
(dashed curve)
phase boundaries change into new ones (solid curves)
in the presence of two competing interactions.
For $V=1.5$ the major change of the phase boundary appears in the low 
doping region of the SC state. Near half-filling, due to 
strong local repulsion, the SDW state is formed throughout the lattice and 
it strongly breaks time-reversal symmetry, leading to the destruction 
of the SC state.
But the AF-only phase remains virtually unchanged at this strength.
In the low doping region a new phase appears where the AF and SC and 
$\pi$-triplet order parameters are non-vanishing,
as described earlier.
For $V=3.0$ where the strengths of AF and SC interactions become comparable,
both the AF-only and SC-only phase boundaries significantly change. 
As in $V=1.5$ the SC phase boundary is most  
strongly modified near half-filling, 
while the AF phase boundary far away from half-filling where AF order is most 
vulnerable to other interactions.
Because of the larger value of $V$, the SC phase region becomes 
wider than that of $V=1.5$.
At a particular point where four different phases (paramagnetic 
metal, SC, AF, and AF+SC+$\pi$-triplet phases) meet,
a tetracritical point\cite{Fisher:1974,Zhang:1997} appears. 
That tetracritical point does not have SO(5) symmetry in general, 
but it has SO(3)$\otimes$U(1) symmetry. SO(5) symmetry is 
realized only when $t'=0$.\cite{Henley:1998}

     In Fig.\ref{fig2} we consider phase diagrams with the next 
nearest neighbor hopping ($t'=-0.3$). This value is close to 
what is suggested for YBa$_2$Cu$_3$O$_{7-\delta}$ and 
Bi$_2$Sr$_2$CaCu$_2$O$_{8+\delta}$.
For small $V$ the coexistence region is more suppressed near half-filling 
than that for $t'=0$ in Fig.\ref{fig1}-(a) because 
the effective density of states for SC is reduced.
For $V=3$, however, the coexistence region is more strongly modified
far away from half-filling 
than in the $t'=0$ case, since the increased effective strength for SC 
in this doping range pushes the AF phase to the left.
The slight increase of the AF region near $x=0.3$ with $t'=-0.3$ is 
presumably due to an artifact of the {\em commensurate} SDW approximation.  

    As in the previous figures, 
the dotted and dashed curves in Fig.\ref{fig3} denote 
the order parameters of AF with $V=0$
and those of SC with $U=0$, respectively.
The order parameters are plotted in Fig.\ref{fig3}-(a) 
as a function of doping ($x$)
with temperature fixed at 0.3. 
As can be inferred from Fig.\ref{fig1}-(b), the order parameters of 
AF and SC states are most strongly affected  
far away from half-filling and near half-filling, respectively,
by the presence of competing interactions.
Where two order parameters (AF, SC) coexist, the third 
$\pi$-triplet order parameter appears without its driving potential ($W=0$).
Its amplitude is maximum when those of AF and SC become comparable
as is implied in Eq.\ref{eq17}
and the maximum value is approximately one fifth of those of 
AF and SC at the corresponding doping level. 
When one of the order parameters appears or disappears,
the other order parameter changes in its slope, a characteristic  
feature of a mean-field approximation.
In Fig.\ref{fig3}-(b)  
the order parameters are plotted as a function of temperature
at fixed doping 0.1. 
Because of their competing nature, both order parameters 
are reduced at low temperatures from those of 
AF-only and SC-only phases.

   In Fig.\ref{fig4} the phase diagram is shown for 
$W=3$, $V=3$, and $t'=0$.
As in the previous figures, the dotted, dashed, and long-dashed curves 
are phase boundaries for AF with $V=W=0$, for SC with $U=W=0$ and 
for $\pi$-triplet with $U=V=0$.
Due to the $W$ term, AF and SC regions are increased compared with those for 
$W=0$ case (Fig.\ref{fig1}-(b)).
This is because the $W$ term increases the stability of AF and SC 
by converting SC (AF) electrons into AF (SC) electrons
instead of destroying them.
One of the interesting features is that a $\pi$-triplet phase does 
not appear as a separate phase for $W=3$, although $W$ is as large as $V$.
In order to create such a 
phase, $W \geq 4.5$ is required.
Then a SC-only phase disappears in the phase diagram and  
the roles of SC and $\pi$-triplet 
are interchanged. In this situation
the SC order parameter appears only where AF and $\pi$-triplet phases 
coexist.

   In the last figure 
the order parameters are plotted as a function of 
doping and temperature for $W=3$, $V=3$, $t'=0$.
As before the dotted, dashed, and long-dashed curves 
are the order parameters for AF with $V=W=0$, for SC with $U=W=0$ and 
for $\pi$-triplet with $U=V=0$, respectively.
Because of the $W$ term, the critical concentration of AF order at $T=0.3$
(Fig.\ref{fig5}-(a)) is increased from $x=0.2$  
for $W=0$ to $x=0.27$, much closer to 
that of AF order
with $V=W=0$ (dotted curve). Thus 
the order parameter of AF with $W=3$, is similar to 
that of AF with $V=W=0$.
The maximum value of $\pi$-triplet order parameter is 0.051, 0.064, and 
0.091 for $W=0$, 1.5, and 3.0, respectively.
This indicates the robustness of dynamically  
generated ($W=0$) part of the $\pi$-triplet 
order parameter.

     Before closing several comments are in order. In order to 
find the nature of phase transition at the phase boundary of 
AF (or SC) $\leftrightarrow$ AF+SC+$\pi$-triplet, we calculated  
the free energies of the three different phases (AF, SC, and 
AF+SC+$\pi$-triplet)
at constant temperature and chemical potential.
Inside the phase boundary the free energy of 
AF+SC+$\pi$-triplet is lowest\cite{Comment0} and its slope does not change 
as the temperature crosses the phase boundary. 
Thus in a mean-field approximation the phase transition of 
AF (or SC) $\leftrightarrow$ AF+SC+$\pi$-triplet is 
second order.\cite{Comment1}
This is consistent with the earlier result of 
Kato and Machida.\cite{Kato:1988} 
These authors classified anisotropic
pairing states enumerated group theoretically into two categories:
less-competing and more-competing states according to the combined  
symmetry of the SC order parameter (parity and translational symmetry
determined by the nesting vector).
Less-competing states yield second order phase transition between 
AF (or SC) and $\leftrightarrow$ AF+SC+$\pi$-triplet, while 
more-competing states yield first order phase transition. 
D-wave SC belongs to the less-competing states according to their 
classification.
In this paper 
only a commensurate magnetic structure is considered 
in the SDW approximation.
From more sophisticate considerations,\cite{Schulz:1990} 
away from half-filling and at low temperatures
the commensurate magnetic order should be replaced by 
an incommensurate one.
Lastly, mean-field theories neglect fluctuations completely, thus they 
are inadequate in low dimensions because there is no long-range 
order at finite temperatures in one and two dimensions.\cite{Mermin:1966}
This pathological feature may be avoided by considering   
a small z-axis component.  
Since strong fluctuations still remain due to quasi two dimensional nature,
the phase boundaries will be shrunk drastically.
For AF state the boundary will be squeezed toward the low temperature  
and low doping, while  
for SC state toward the low temperature.
Accordingly the coexistence region may remain in a very 
small region of the phase diagram or even disappear.\cite{Comment2}
Then the mean-field critical temperatures will be replaced by a
crossover temperature $T^{*}$ below which 
a pseudogap\cite{Vilk:1997} will appear in the single particle   
spectral function.  
This precursor effect of the ground state can be caused by antiferromagnetic  
or superconducting fluctuations depending on the location in the phase 
diagram.
Likewise, 
a strongly fluctuating 
$\pi$-triplet amplitude will manifest itself 
in the region surrounded by the AF and SC mean-field phase boundaries.

     In conclusion,
we have studied 
the close interplay between antiferromagnetism and d-wave superconductivity
in a mean-field approximation.
In the presence of competing interactions the phase boundaries  
of AF and SC states
are significantly modified in some region of the doping-temperature
plane.
In the coexistence region of AF and d-wave SC
a new order of spin-triplet pair is dynamically generated without 
its driving interaction.
We argue that 
this feature is not restricted only to the current problem,
but is a generic feature for fermionic systems.
The dynamical generation of 
spin-triplet order parameter is found to be robust
to variations in the mean-field Hamiltonian.
Fluctuation effects to the mean-field result are briefly discussed.

   The author would like to thank Prof. Tremblay for numerous help 
in various stages of this work as well as for critical reading of the
paper.
He also thanks Prof. S\'{e}n\'{e}chal for helpful discussions
in the early stage.
The present work was supported by a grant from the Natural Sciences and
Engineering Research Council (NSERC) of Canada and the Fonds pour la
formation de Chercheurs et d'Aide \`a la Recherche (FCAR) of the Qu\'ebec
government.
\appendix 
\section*{Transformation matrix}

    The transformation matrix $A_{i,j}$
relating the original $c$ operator
to the eigen-operator $\gamma$ is defined by 
\begin{eqnarray}
\psi_{i}(\vec{k})=\sum_{j}A_{i,j}\gamma_{j}(\vec{k}) \; .
                                                           \label{eq18}
\end{eqnarray}
The components of the first column vector are given as 
\begin{eqnarray}
& A_{1,1} & =((E_{+}(\vec{k})+\varepsilon_{\vec{k}})
         (E_{+}(\vec{k})+\varepsilon_{\vec{k}+\vec{Q}})
         (E_{+}(\vec{k})-\varepsilon_{\vec{k}+\vec{Q}})
                                             \nonumber  \\
& &     -(E_{+}(\vec{k})-\varepsilon_{\vec{k}+\vec{Q}})(Um)^{2}
        -(E_{+}(\vec{k})+\varepsilon_{\vec{k}})(Vs\phi(\vec{k}))^{2}
                                             \nonumber  \\
& &     -(E_{+}(\vec{k})+\varepsilon_{\vec{k}+\vec{Q}})(Wt\phi(\vec{k}))^{2}
        +2(Um)(VS\phi(\vec{k}))
                                             \nonumber  \\
& &      (Wt\phi(\vec{k})))
         /\sqrt{\sum_{i} A^{2}_{i,1}}
                                             \nonumber  \\
& A_{2,1} & =((VS\phi(\vec{k}))[
         (E_{+}(\vec{k})+\varepsilon_{\vec{k}+\vec{Q}})
         (E_{+}(\vec{k})-\varepsilon_{\vec{k}+\vec{Q}})
                                             \nonumber  \\
& &     -(Um)^{2}
        -(VS\phi(\vec{k}))^{2}
        +(Wt\phi(\vec{k}))^{2}]
                                             \nonumber  \\
& &     +2(Um)(Wt\phi(\vec{k}))
         \varepsilon_{\vec{k}+\vec{Q}})
        /\sqrt{\sum_{i} A^{2}_{i,1}}
                                             \nonumber  \\
& A_{3,1} & =-((Um)[
         (E_{+}(\vec{k})+\varepsilon_{\vec{k}})
         (E_{+}(\vec{k})+\varepsilon_{\vec{k}+\vec{Q}})
                                             \nonumber  \\
& &     -(Um)^{2}
        -(VS\phi(\vec{k}))^{2}
        -(Wt\phi(\vec{k}))^{2}]
                                             \nonumber  \\
& &     +(Vs\phi(\vec{k}))(Wt\phi(\vec{k}))
         (2E_{+}(\vec{k})
         +\varepsilon_{\vec{k}+\vec{Q}}
         +\varepsilon_{\vec{k}}))
                                             \nonumber  \\
& &        /\sqrt{\sum_{i} A^{2}_{i,1}}
                                             \nonumber  \\
& A_{4,1} & =((Wt\phi(\vec{k}))[
         (E_{+}(\vec{k})+\varepsilon_{\vec{k}})
         (E_{+}(\vec{k})-\varepsilon_{\vec{k}+\vec{Q}})
                                             \nonumber  \\
& &     -(Um)^{2}
        +(VS\phi(\vec{k}))^{2}
        -(Wt\phi(\vec{k}))^{2}]
                                             \nonumber  \\
& &     +(Um)(Vs\phi(\vec{k}))
         (\varepsilon_{\vec{k}+\vec{Q}}
         +\varepsilon_{\vec{k}}))
        /\sqrt{\sum_{i} A^{2}_{i,1}} \; ,
                                                           \label{eq19}
\end{eqnarray}
where the square of normalization of the vector is  
\begin{eqnarray}
& & \sum_{i} A^{2}_{i,1} =
2 E_{+}(\vec{k})g(\vec{k}) \Bigl[
 (E_{+}(\vec{k})+\varepsilon_{\vec{k}})
                                             \nonumber  \\
& & (2g(\vec{k})+\varepsilon^{2}_{\vec{k}}-\varepsilon^{2}_{\vec{k}+\vec{Q}})
 +2(Um)^{2}(\varepsilon_{\vec{k}}+\varepsilon_{\vec{k}+\vec{Q}})
                                             \nonumber  \\
& & +4(Um)(VS\phi(\vec{k}))(Wt\phi(\vec{k}))
 +2(Wt\phi(\vec{k}))^{2}(\varepsilon_{\vec{k}}-\varepsilon_{\vec{k}+\vec{Q}})
                  \Bigr]  \; . 
                                             \nonumber  \\
                                                           \label{eq20}
\end{eqnarray}
By symmetry the second column vector $A_{i,2}$ is obtained from 
the first one as follows:
$A_{1,2}=-A_{2,1}, A_{2,2}=A_{1,1}, A_{3,2}=A_{4,1}, A_{4,2}=-A_{3,1}$.
The third column vector may be expressed in terms of 
the first one:
$A_{1,3}=A_{3,1}, A_{2,3}=-A_{4,1}, A_{3,3}=A_{1,1}, A_{4,3}=-A_{2,1}$
with the following replacement in $A_{i,1}$,
$ \varepsilon_{\vec{k}} \leftrightarrow \varepsilon_{\vec{k}+\vec{Q}},
g(\vec{k}) \leftrightarrow -g(\vec{k}), 
E_{+}(\vec{k})
\leftrightarrow E_{-}(\vec{k})$.
Finally the fourth column vector $A_{i,4}$ is found from the  
third one by symmetry
$A_{1,4}=A_{2,3}, A_{2,4}=-A_{1,3}, A_{3,4}=-A_{4,3}, A_{4,4}=A_{3,3}$.
\newpage
\begin{figure}
 \vbox to 7.0cm {\vss\hbox to -5.0cm
 {\hss\
       {\includegraphics{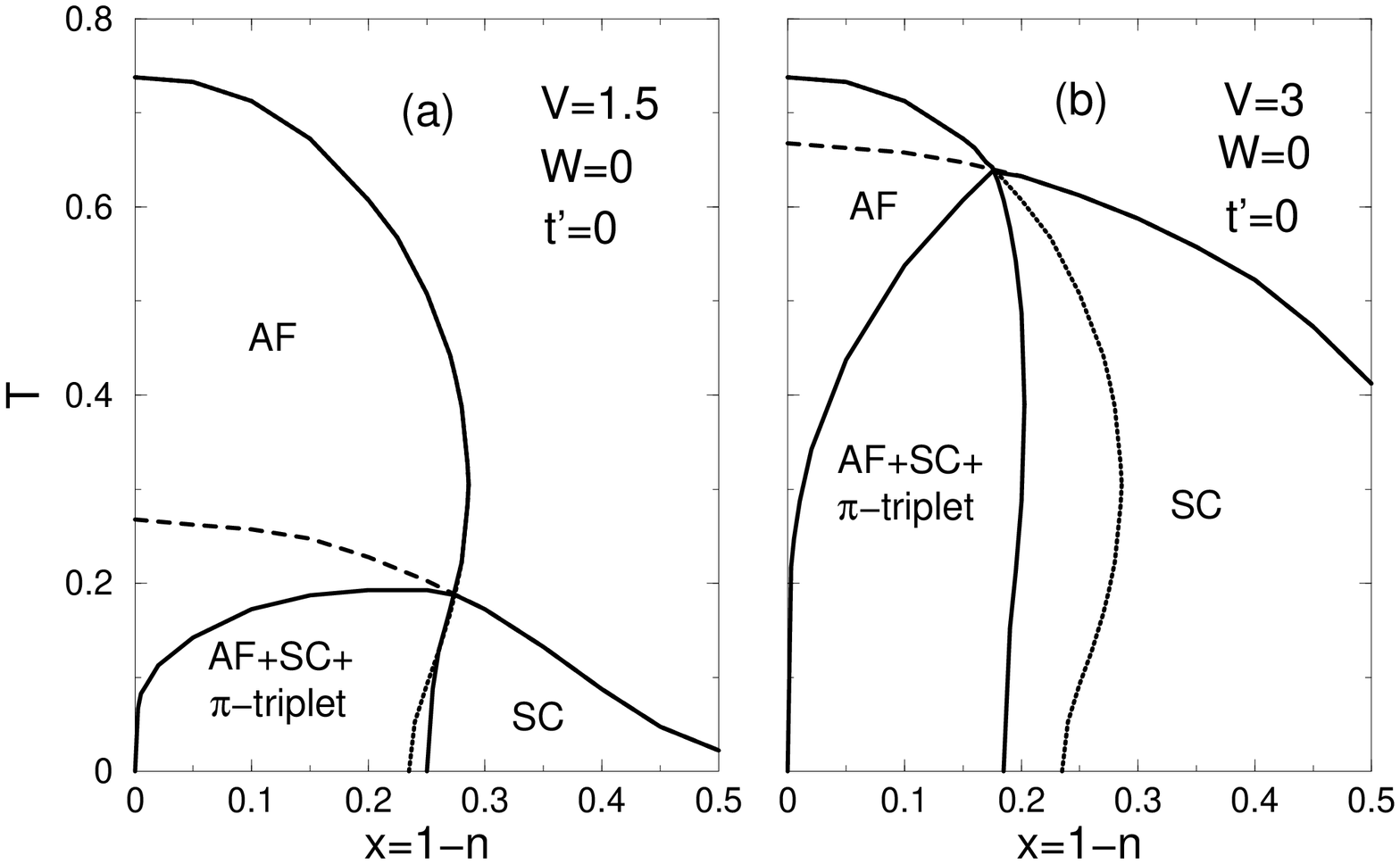}
       }
  \hss}
 }
\caption{Phase diagrams in doping ($x=1-n$) and temperature ($T$) plane 
         for (a) $V=1.5$, and (b) $V=3$, with $U=4$, $W=0$, and $t'=0$.
         The dotted and dashed curves denote the phase boundaries of AF 
         with $V=0$ and of SC with $U=0$, respectively.} 
\label{fig1}
\end{figure}
\begin{figure}
 \vbox to 7.0cm {\vss\hbox to -5.0cm
 {\hss\
       {\includegraphics{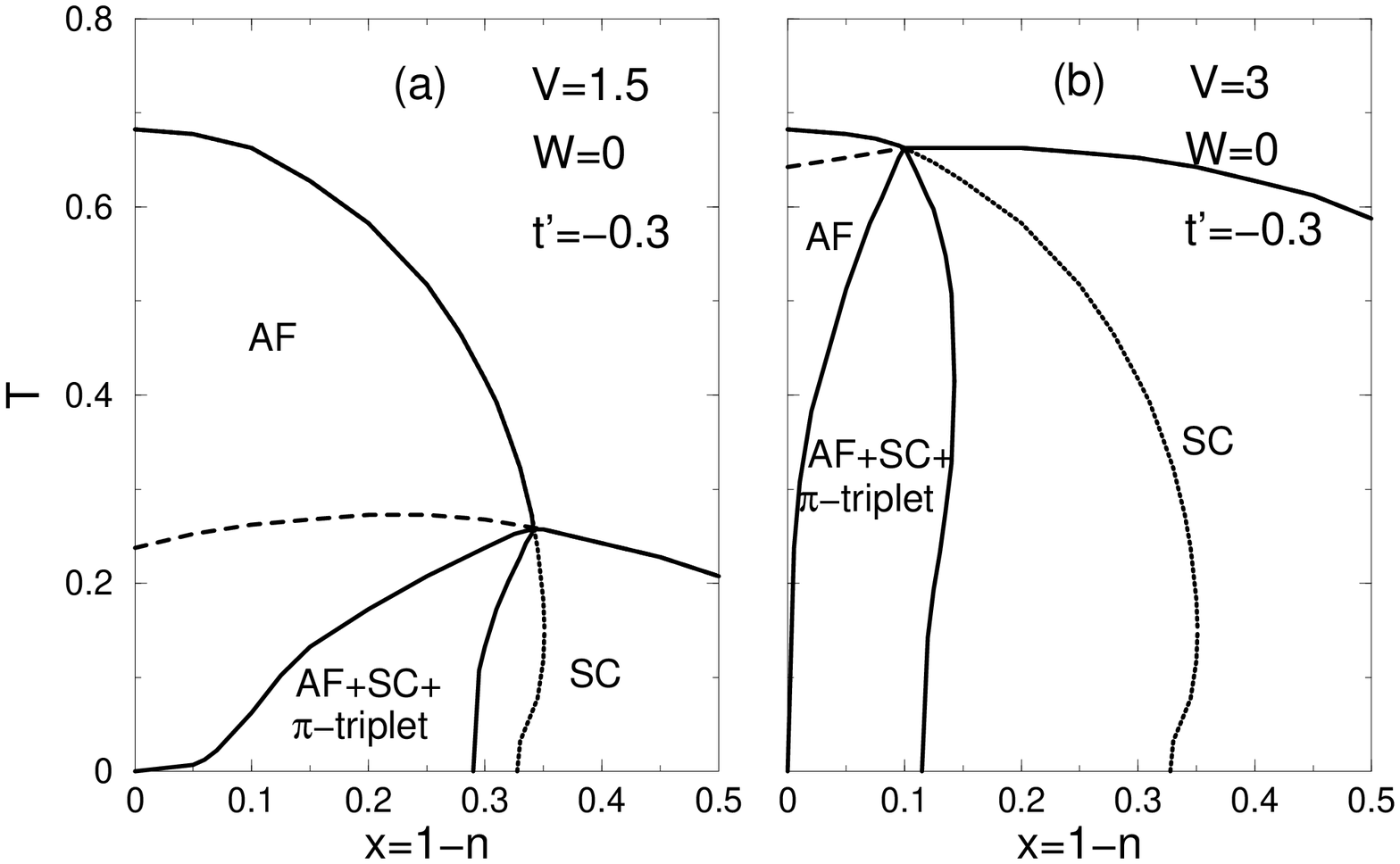}
       }
  \hss}
 }
\caption{Phase diagrams in doping ($x=1-n$) and temperature ($T$) plane 
         for (a) $V=1.5$, and (b) $V=3$, with $U=4$, $W=0$, and $t'=-0.3$.
         The dotted and dashed curves denote the phase boundaries of AF 
         with $V=0$ and of SC with $U=0$, respectively.} 
\label{fig2}
\end{figure}
\begin{figure}
 \vbox to 7.0cm {\vss\hbox to -5.0cm
 {\hss\
       {\includegraphics{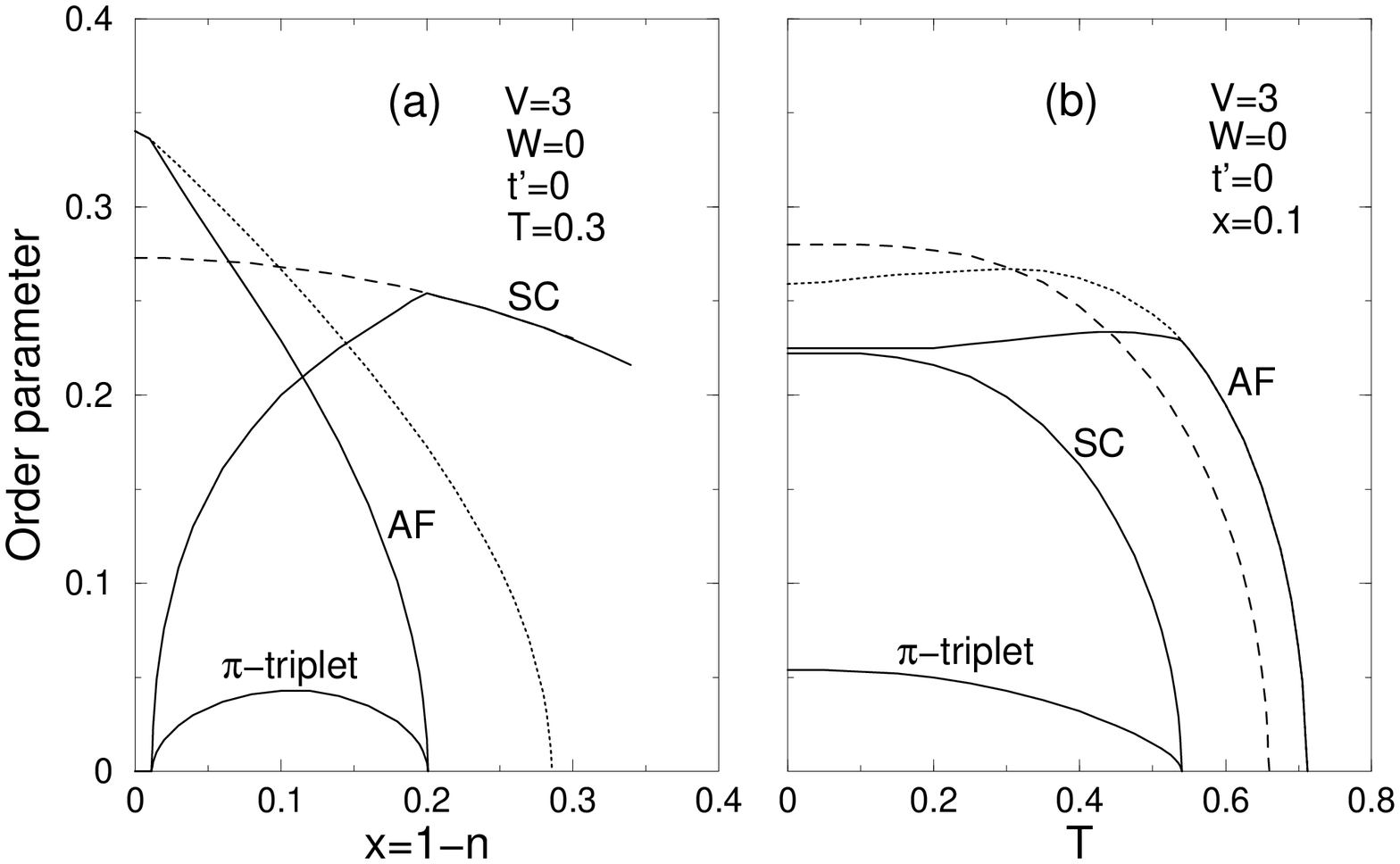}
       }
  \hss}
 }
\caption{Order parameters 
        at (a) $T=0.3$ and (b) $x=0.1$, for $U=4$, $V=3$, $W=0$, and $t'=0$.
        The dotted and dashed curves denote the order parameters of AF 
        with $V=0$ and of SC with $U=0$, respectively.} 
\label{fig3}
\end{figure}
\begin{figure}
 \vbox to 7.0cm {\vss\hbox to -5.0cm
 {\hss\
       {\includegraphics{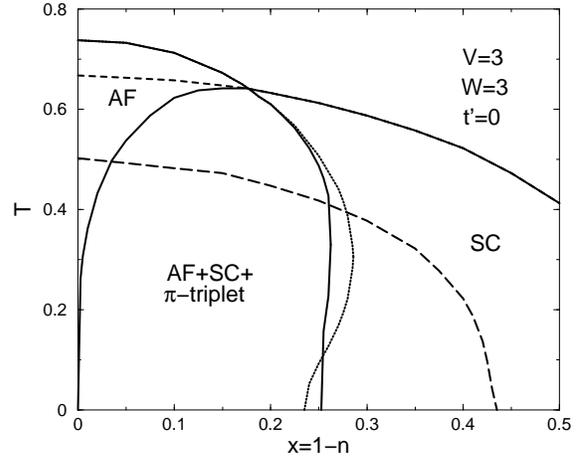}
       }
  \hss}
 }
\caption{Phase diagram in doping ($x=1-n$) and temperature ($T$) plane 
         for $U=4$, $V=3$, $W=3$, and $t'=0$.
         The dotted, dashed, and long-dashed curves denote  
         the phase boundaries of AF 
         with $V=W=0$, of SC with $U=W=0$, of $\pi$-triplet with  
         $U=V=0$,
         respectively.} 
\label{fig4}
\end{figure}
\begin{figure}
 \vbox to 7.0cm {\vss\hbox to -5.0cm
 {\hss\
       {\includegraphics{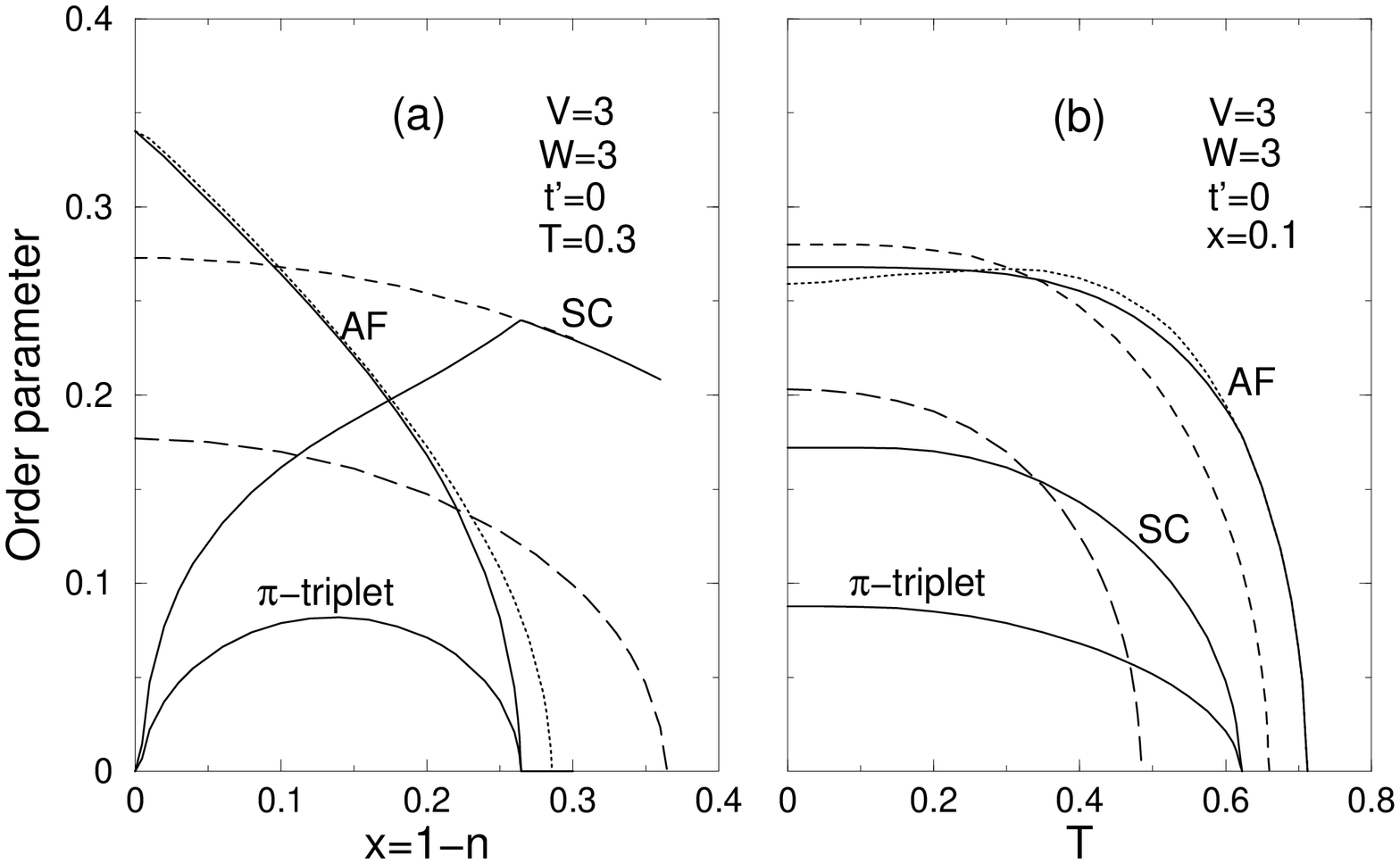}
       }
  \hss}
 }
\caption{Order parameters 
        at (a) $T=0.3$ and (b) $x=0.1$, for $U=4$, $V=3$, $W=3$, and $t'=0$.
        The dotted, dashed, and long-dashed curves denote  
        the order parameters of AF 
        with $V=W=0$, of SC with $U=W=0$, of $\pi$-triplet with  
        $U=V=0$,
        respectively.} 
\label{fig5}
\end{figure}

\begin{references}
\bibitem{Comment1} Note that the AF+SC+$\pi$-triplet phase transforms
         as a subgroup of either the AF or the SC phases, as required 
         from general considerations.
\bibitem{Comment2} Another mechanism that could lead to the disappearance 
         of the AF+SC+$\pi$-triplet phase is disorder. Indeed, the 
         triplet component of the mixed phase should be very sensitive 
         to disorder.
\end{references}
\end{document}